# Polymer architecture orchestrates the segregation and spatio-temporal organization of replicating *E. coli* chromosomes in slow growth.

Debarshi Mitra, Shreerang Pande & Apratim Chatterji*

*Dept of Physics, IISER-Pune, Pune, India-411008.*

(Dated: December 23, 2021)

The mechanism and driving forces of chromosome segregation in the bacterial cell cycle of *E. coli* is one of the least understood events in its life cycle. Using principles of entropic repulsion between polymer loops confined in a cylinder, we use Monte carlo simulations to show that the segregation dynamics is spontaneously enhanced by the adoption of a certain DNA-polymer architecture as replication progresses. Secondly, the chosen polymer-topology ensures its self-organization along the cell axis while segregation is in progress, such that various chromosomal loci get spatially localized. The time evolution of loci positions quantitatively match the corresponding experimentally reported results, including observation of the cohesion time and the ter-transition. Additionally, the contact map generated using our bead-spring model reproduces the four macro-domains of the experimental Hi-C maps. Lastly, the proposed mechanism reproduces the observed universal dynamics as the sister loci separate during segregation. It was already hypothesized and expected that SMC proteins, e.g. MukBEF contribute over and above entropic repulsion between bacterial-DNA ring-polymers to aid the segregation of daughter DNAs in the *E.coli* cell cycle. We propose that cross-links (plausibly induced by SMC proteins) at crucial positions along the contour is enough to provide sufficient forces for segregation within reasonable time scales. A mapping between Monte Carlo diffusive dynamics time scales and real time units helps us use experimentally relevant numbers for our modeling.

**Teaser:** *The mechanism governing the separation of two sister chromosomes into two halves of the cell be- fore cell division is known for higher organisms, but not yet fully understood for bacterial cells which are much simpler organisms. For E.coli bacterial cells, it is es- tablished that certain chromosomal segments get localized along the cell long-axis while replication is in progress. Our simulations establish that a specific looped architec- ture for the replicating chromosomes provides sufficient entropy driven repulsive forces between loops leading to localization of chromosomal segments and their segrega- tion within observed time scales. Our proposed frame- work reconciles many spatio-temporal organizational as- pects of E. coli chromosome as seen in-vivo and provides a consistent mechanistic understanding of the process un- derlying segregation.*

## I. INTRODUCTION

The replication and segregation of chromosomes into two  halves of the cell before the start of cell division is an important step to maintain the life cycle of a liv- ing cell [1, 2].   Nature has devised advanced strategies to segregate the chromosomes of eukaryotes, but inter- estingly,  the mechanism of chromosome segregation of a much simpler single-celled bacterial organism such as the *E. coli* is unknown till date [3–8]. Investigators have identified mechanisms with contributions from proteins which leads to the chromosomal segregation in bacteria *B. subtilis* [8, 9] and *C. crescentus* [3, 10, 11], but any such mechanism could not be identified for *E. coli*.

* Correspondence email address:    apratim@iiserpune.ac.in

The chromosome of *E. coli* can be thought of as hav- ing ring polymer architecture, and it has been proposed that the physical principles related to entropy maximiza- tion (or equivalently free energy minimization) might be used by nature to achieve chromosome segregation while replication is in progress. It has been shown that en- tropic repulsion between 2 ring polymers confined within a cylinder lead to their gradual segregation [12–18]. But then it has to be aided by some other hitherto unknown processes driven by proteins, as entropic repulsion is too weak [8] to have segregation within time scale of 60 min- utes of the C+D-period [19–21]. This has spawned a series of experiments over the last two decades.

Detailed in-vivo experiments, with DNA-loci labelled with fluorescent markers, have helped us visualize their spatial dynamics as the *E. coli* DNA replicates and simul- taneously segregates. The DNA-organization transitions from the ori-ori-ter configuration to the ori-ter-ori con- figuration in the midst of segregation [22, 23]. As repli- cation proceeds, the rod shaped cell elongates along its long-axis to double its length before cell division occurs. Furthermore,  for slow growth conditions (SGC) there is a cohesion time of around 15 minutes (post replication), before the two *oriCs* branch out from the center of the  cell and get localized at the two quarter positions along the cell long axis [23]. The variation of *oriC*-position, as measured by the standard deviation of its mean position, is less than one tenth of the length of the cell. Upto 7 other loci are found to be well localized at certain po- sitions along the cell long-axis [23]. In fast growth con- ditions, with multi-fork replication, the *oriCs* are found  in positions (1/8, 3/8), (5/8, 7/8) position along the cell long axis. Though the mechanism of the localization of the *oriC* and other loci remain unexplained, researchers have observed the increase in MukBEF at these positions during segregation. It has been suggested that the positions of the *oriC*s are correlated with the increase of Muk-BEF concentrations [24]. Other dynamical aspects of the segregation process remain unexplained (or controversial) as well, e.g., (*i*) the sequential start of segregation of the different loci along the chain contour from the cell center (and if that is consistent with the train-track model or the replication factory model) and (*ii*) the origin of the universal values of drift velocity ($0.33\mu/min$) of loci at the onset of segregation, observed for 8 pairs of genetic loci located on the two daughter chromosomes, respec- tively [23]. The daughter chromosomes remain conjoined at the *dif* position in the ter domain near the cell cen- ter till the end of replication, and move apart thereafter.

The experiments, which mark the spatio-temporal distribution of the DNA molecule in individual cells during the segregation process, however lack sufficient resolution. One typically follows just 8−13 loci on the DNA molecule. Complementary experiments, which provide more detailed global organization of the DNA-polymer by 5C and Hi-C contact maps [25–28], lose out on the temporal information. These show the existence of topologically associated domains (TADs) and point to the existence of 4 macro-domains in the DNA-organization in the *E. coli* cell [29–32]. The positional correlations presented in Hi-C maps is from data collected from a population of many *E. coli* cells which are at different stages of the cell cycle. Since the replication and segregation constitute of 80% of the cell cycle, any model of segregation should also be consistent with data presented in Hi-C maps. Researchers have several theoretical models (e.g. the loop extrusion and the SBS model) to study the organisation and segregation of chromosomes, both eukaryotic and prokaryotic, which are optimised to obtain a match with the experimental Hi-C map [33–40]. Consensus has not been reached about the origin of the four macro-domains in *E.coli*, though SMC proteins are expected to play a important role [32, 36, 41].

Many studies using coarse-grained polymer models have primarily focused on the organization of the chromosome in bacterial or eukaryotic cells [42–53] but relatively fewer on the segregation dynamics of chromosomal loci while replication is in progress [24, 54, 55]. A more detailed framework which reconciles complementary experimental results pertaining to both the spatial and temporal organization of bacterial chromosomes during its life cycle in slow [23] and fast growth conditions [22], incorporating the reported effects of SMC proteins in vivo, is in order [56]. This work aspires to propose an unifying framework addressing these concerns for a replicating chromosome in slow growth conditions.

Our proposed model relies on the concepts of enhanced entropic repulsion provided by the presence of loops within each of the two daughter DNA-polymers and entropic repulsion between loops of the same daughter chromosome. These loops are created by two additional contacts/links which bridge distant chromosomal regions: cross-links (CLs) created between specific pair of points along the chain contour in the polymer physics terminology. Thereby, we propose that the daughter DNAs adopt a slightly more evolved polymer architecture than just a simple ring polymer, as replication proceeds. This simple but pivotal idea provides a consistent picture of the segregation of *E. coli* daughter DNAs after replication, and quantitatively integrates disparate experimental measurements on (a) localization of loci along cell long-axis, (b) observation of four macro-domains in Hi-C maps, (c) universal speeds of loci at the onset of segregation before reaching their home positions and (d) enhanced propensity to segregate resulting in lower segregation time scales of daughter DNAs.

The polymer architectures 'Arc-2' and 'Arc-2-2' that we propose subsequently in this manuscript, has been devised in particular to reproduce the specific localisation of *oriC*s along the cell long axis, and the macrodomain organisation of *E.coli* chromosomes. To this end we introduce loops to induce entropic repulsion between different polymer segments within a daughter chromosome. This also leads to enhanced entropic repulsion between newly replicated sister chromosomes. This intra-chromosomal entropic repulsion successfully leads to the localisation of *oriC* to the mid-cell position before replication and to the quarter positions after replication. Additionally it also leads to the localisation of other loci as seen *in-vivo*. We discuss 'Arc-2' in detail although it is not the underlying architecture for *E.coli* chromosomes since it is incompatible with the Hi-C data. However, insights gained from the 'Arc-2' architecture led us to the 'Arc-2-2' architecture which is likely adopted by the *E.coli* chromosome and is in agreement with the observation of macrodomains from Hi-C data for *E.coli* chromosomes.

Nucleoid associated proteins (NAPs) play an important role in *E.coli* segregation [14, 32]. They compact- ify the bacterial DNA at all length scales and also help bridge distant chromosomal segments. It has been shown previously that depletion of some SMC proteins like Muk-BEF leads to defects in the segregation of the *E.coli* chromosome, thus underlining their importance [25, 32, 57]. We propose that the role of SMCs (and MukBEF in particular) is to facilitate the segregation of replicating chromosomes by modifying their loop architecture suitably, such that they are more amenable to segregation by entropic means. Furthermore previous studies already suggest that the *E.coli* chromosome is organized into 'loop domains' [58–60]. Here we identify particular architectures, which increase segregation rates to bring down the segregation time and more importantly leads to the localisation patterns of chromosomal loci, as seen in-vivo. We achieve this by employing a modified ring-polymer architecture of the chromosome, which leads to entropic repulsion of both 'intra' and 'inter' chromosomal polymer segments. It is already known that a branched/loop topology expedites the segregation process [13, 14]. However a causal link between polymer topology and localisation of chromosomal loci has not been established till date. Several previous studies have adopted the route of optimising Hi-C contact maps to extract simulation parameters and studying the resultant organisation [85, 61]. We however take an alternate approach and develop a minimal polymer based framework. Such a simplified representation in our opinion helps us to unearth key insights pertaining to the role of loops in the entropic organisation of the chromosome, which might otherwise be missed with more detailed models.

The unreplicated DNA of *E. coli* having 4.6 million bp (base pairs), is represented as a bead-spring ring polymer with 500 beads, modified with additional CLs, also refer Fig. 1. Thus each bead represents around 9200 bp. It is known that that there are topologically distinct domains of size $10 kbp$ in *E.coli* due to gene expression [32, 62]. We also expect contributions from various histone-like HU-proteins (and other NAPs) to compactify the DNA at various length scales. But we expect, that the two cross-links at specific sites considered for this study to be relatively long-lived, while the rest get on to the backbone or fall off at different times to complete other biological functions, during the life cycle of the cell. These specific CLs bridge segments which are at a distance about $\sim 1 Mbp$ away from each other to create the Arc-2 architecture. The CLs may arise due to MukBEF-activity which are known to create contacts between segments separated by $\sim 900 kbp$ [32, 63]. Two *additional* CLs at specific sites in the *ter*-macrodomain bridge segments which are 600 kbp apart; we name this as the Arc-2-2 architecture. Thus the CLs are chosen such that the segments in the ter macrodomain, are linked to relatively closer segments. Our choice of specific sites for CLs in both Arc-2 and Arc-2-2 was found to be consistent with the recently reported observation that MukBEF activity is hindered in the *ter*-macrodomain due to the presence of MatP [32], which possibly leads to the formation of smaller loops in the ter-macrodomain. Furthermore, MatP is known to condense the *ter* macrodomain by inducing intra-chain specific interactions. Thus in Arc-2-2 we introduce two smaller loops (in the *ter* domain) in addition to those of Arc-2 architecture, as we show later. The choice of the specific cross-linking sites reported here have been chosen from a pool of several other possible polymer-loop architectures having different CL-sites. We use insights gained from our other studies (reported elsewhere [64]) to focus on the consequences of *E.coli* adopting (Arc-2 and) Arc-2-2 architecture(s) due its higher degree of success in reconciling the experimental results pertaining to localisation of chromosomal loci and speed of segregation.

**Model** We perform Monte Carlo (MC) simulations starting with a polymer of Arc-2 or Arc-2-2 architecture. The unit of length in our simulations is $a$, which is the equilibrium length of springs between adjacent monomers along chain contour. Each monomer is of diameter $\sigma = 0.8a$ and excluded volume interactions are implemented by the WCA potential. The first monomer in the ring is labelled as the *oriC* and monomer number 250 is assumed to be the *dif*-locus in *ter* domain, here-after referred to as *dif-ter*. If we chose $a \approx 150$ nm, which we justify to be a reasonable length scale for 9200 base pairs considering DNA-compaction at smaller length scales, then we can establish that $25 \times 10^3$ Monte Carlo steps (MCS) corresponds to 1 second(s) of real time: refer SI-1 and Fig.S1 in the Supplementary Information (SI).

The complex machinery of opening of the mother chromosomal strands by helicase and creating a daughter ds-DNA from the resulting single strands is significantly simplified in our model. We model replication explicitly, by adding two monomers (representing 9200bp each) at discrete replication steps (every $2 \times 10^5$ MC steps, i.e., every $\approx 8$ seconds) at a location proximal to the two RFs, to create another chain labelled DNA-2. The position of two replication forks (RFs) is assumed to start moving bidirectionally from one monomer to the next starting from the *oriC* towards *dif-ter* as the simulation proceeds. This ensures the correct replication rates; also refer SI-II and Fig.S2 for details. After each replication step the RF position moves ahead to the next monomer, thereafter, the original (mother) strand from *oriC* to the RF position is re-labelled as daughter DNA-1. To mimic the effect of topoisomerase which allows release of topological constraints by facilitating chain crossing, we reduce the $\sigma$ to $\sigma' = 0.1a$ at regular intervals for nearly 8% of the total evolved time: refer SI-II.

We introduce cross-links (CLs) between monomer 1 and 125 immediately after 125-th monomer is replicated in strand DNA-2, also refer schematics in Fig. 1. Simultaneously, we also cross-link monomer 1 to 375. Note that the 375-th monomer gets replicated at the same time as the 125-th monomer, refer Fig. 1. These additional CLs introduced in daughter DNA-polymer DNA-2 are realised by harmonic springs of length $a$. We call this the Arc-2 architecture. The introduction of the CLs leads to the subsequent movement of the monomer pairs linked by CLs towards each other, unless prevented by topological constraints, which can be released subsequently. We appreciate that the directed movement of crosslinked monomers towards each other, as modelled by us is unlikely to occur *in-vivo*. In reality the spatial proximity of genomic segments is likely to be realised by other physical mechanisms before they get cross-linked. These mechanisms may include that of liquid-liquid phase separation (LLPS), which may lead to aggregation of specific genomic segments [65, 66]. However in our coarse grained simulation model we do not take such microscopic details into account. Instead we investigate the consequences of the bacterial chromosome-polymers adopting a polymer architecture due to the aforementioned cross-links.Furthermore it is quite plausible that the process of "cross-linking" which we have implemented in the simulations, might be realised by distant genomic segments diffusing closer to each other. Thus it is possible that the "effective crosslinks" proposed herein, is achieved in reality over longer timescales than those implemented here. Future experimental studies may shed light on this aspect.

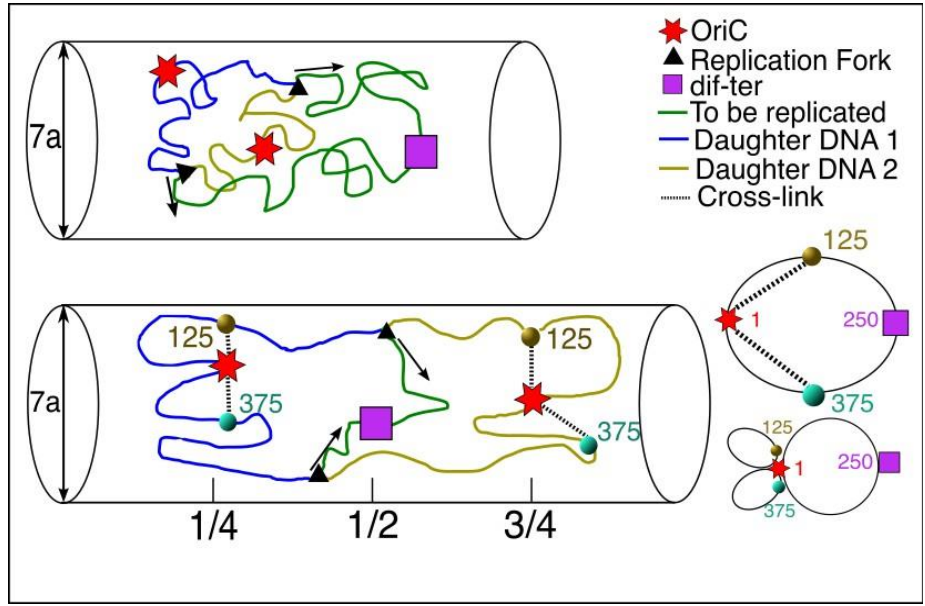

Figure 1. **Schematic showing simulation model and polymer architecture Arc-2**. The cylinder at the top shows the *ori-ori-ter* DNA-polymer configuration soon after the onset of replication. Just after replication begins at the *oriC*, there are two *oriC* loci as shown in the schematic. The chain contour(s) joining the two *oriC*-s to the two replication forks (RFs) are the two daughter DNAs, labelled as daughter DNA-1 (blue) and Daughter DNA-2 (light green), respectively. The two replication forks (RFs) are placed on the two arms of the mother DNA (to be replicated DNA) shown in deep green, and form the junctions between DNA-1, DNA-2 and the mother DNA. In our model the light green chain containing the *oriC* is considered to be Daughter DNA-2, and is made by adding monomers explicitly, proximal to the RFs at the replication step every 2 $10^5$ MCS. Thereafter, the RFs move to the neighbouring monomer towards *dif-ter*. As the RFs move along the chain, the chain length of DNA-1 (blue) and DNA-2 (light green) increases and the length of the chain corresponding to the mother DNA decreases. After each replication step, the monomers of mother DNA are re-labelled as Daughter DNA-1 (blue) after the RFs move to the subsequent monomer along the chain, as they approach *dif-ter* bi-directionally. As replication progresses, the configuration spontaneously evolves to the *ori-ter-ori* configuration as seen in the bottom cylinder. The cylinder elongates as replication proceeds. We introduce 2 cross-links at specific segments by connecting specific monomers by the spring harmonic potential, for DNA-2 and thereby create polymer architecture Arc-2, having internal loops within each ring polymer. We start out with the mother DNA having the Arc-2 architecture, and since monomers of mother DNA are re-labelled as DNA-1, it maintains the Arc-2 architecture. We also show a schematic which focuses on the polymer-architecture and clearly identifies the cross-linked monomers in DNA-1 & DNA-2.

At time $t = 0$, a Arc-2 unreplicated DNA-polymer is confined within a cylinder of diameter $D_c = 7a \approx 1\mu$) and length $L_i = 17.5a$, with an aspect ratio of 1 : 2.5 and a colloidal volume fraction of 0.2, as is seen in-vivo. As the the DNA gets replicated, the length of the cylinder doubles itself to be $L_f = 35a$, through an incremental increase every $2 \times 10^5$ MCS. We perform 150 million MCS in one run, which corresponds to ~100 minutes. We do not model the cell division process in our minimal model, which occurs at ~60 minutes in-vivo. Instead, after the completion of replication, the two model daughter chromosomes remain attached at *dif-ter* loci. We follow the positions of different loci of the daughter chromosomes, viz, monomers indexed as 22, 141, 443, 369, 304 and 228 in our ring polymer from the time of replication of these monomers. These position of these monomers along the contour correspond to loci labelled as L1, L2, R1, R2, R3 and *ter1* in [23].

## II. RESULTS

**Dynamics of chromosome segregation:** In Figure 2a, we show a representative snapshot of our simulations at the end of replication, with two daughter DNAs where we can identify the positions of the two *oriC* (red circles) and *dif-ter* (blue cicle). The color gradient indicates the position of a monomer along the chain contour, and the color code is identical for the two DNA-polymers. Thus the reader can see the positional organization of the two polymers as one moves from monomer-1 to 500 along chain contour. Monomers 125 and 375 are close to the *oriC* due to the presence of CLs. We also provide a simulation video, where one can observe the segregation of the daughter chromosomes even as replication is in progress, refer section-Movie in SI. Furthermore, the two *oriC*s (shown in red) and the *dif-ter* locus (shown in yellow) localise around their respective 'home positions'. We note that the configuration changes spontaneously from the ori-ori-ter configuration to the ori-ter-ori con-figuration as the simulation proceeds, which is as seen in-vivo. In Fig.2b, the data shows the distance between the center of mass ($D_{com}$) of the two daughter DNA-polymers for 4 independent runs. We see that segregation is typically complete in $\approx 6 \times 10^7$ iterations. We intentionally also show data for run4, to emphasize the infrequent cases where segregation remains incomplete.

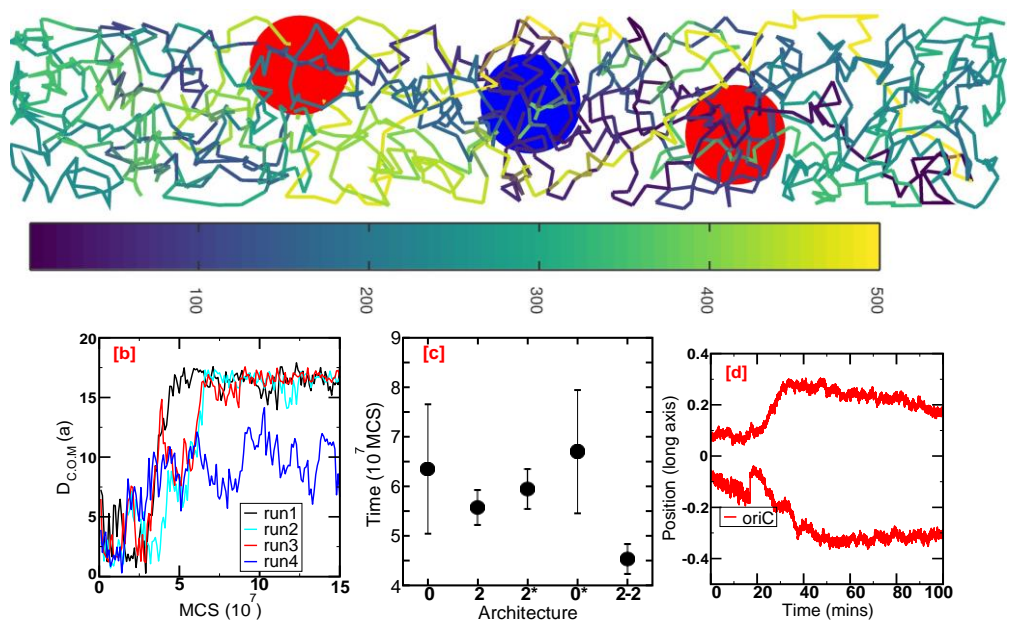

**Figure 2. Dynamics of chromosome-polymer segregation** (a) A representative snapshot of the segregated polymers from our simulation at the end of the replication process at $5 \times 10^7$ MCS (equivalently 33.3 mins) is shown. The length of the cylinder (not shown) is $35a$, diameter remains $7a$ ($a \approx 150nm$). The two red circles represent the *oriC* at the two quarter positions and the blue circle is the *difter* locus. The colour changes as we move along the contour from monomer 1 to 500: refer colour bar. (b) The distance $D_{COM}$ between the two centers of mass (COM) of each daughter DNA is plotted versus MCS, from the start of replication process. We continue till $15 \times 10^7$ MCS, though we expect cell division (not modelled) to occur after $9 \times 10^7$ MCS. Data for 4 independent runs ( cell cycles for 4 individual cells) are shown. (c) The segregation time calculated for different architectures (refer main text) is shown. The segregation times have been computed by averaging over 10 independent runs for each architecture. (d) The position of the 2 *oriCs* are shown as a function of MCS, averaged over 20 independent runs. The cylinder doubles its length from $L_i = 17.5a$ to $L_f = 35a$ by $5 \times 10^7$ MCS, and the *oriC* positions shown are normalized by $L_f$. One can map 25 kMCS to $1s$: refer SI-1, or equivalently $10^7$ MCS $\equiv$ 6mins and 40s.

In Fig.2c, we also compare entropic segregation of two overlapping ring-polymers without loops (Arc-0 architecture with 500 monomers and replication incorporated in modeling) and the polymer architecture with 2 additional loops (Arc-2). When $D_{com}$ exceeds $11a$ ( $1.5D_c$) and that condition is maintained for the next $4 \times 10^7 \times$MCS we deem the daughter chromosomes to have been segregated. We observe faster segregation in 50 million MCS ($\approx 33$ mins) with Arc-2 and even smaller times with Arc-2-2, refer Fig.2(c) and SI-III for other details of calculation. The mechanism of segregation is purely entropic in all the cases. If the two ring-DNA-polymers (Arc-0) occupy different halves of the cylinder they can explore more phase space conformations, and therefore segregate [13]. In the case of architecture Arc-2 (or Arc-2-2), the internal loops of both DNA-1 and DNA-2, in turn repel each other (intra-chromosomal entropic repulsion) as well as the 'internal' loops belonging to the the other daughter DNA (inter-chromosomal entropic repulsion). This 'internally' looped architecture of the chromosomes hastens the segregation process, and as we show later also leads to the localisation of genomic loci. Note that when the daughter chromosomes are *completely* segregated once $D_{com} \nleq L_f / 2 \equiv 17.5a$, which might take a bit longer. We avoid this criteria for identifying segregation, because fluctuations in $D_{COM}$ often does not maintain $D_{COM} > 17.5a$ for $4 \times 10^7$ MCS.

Moreover, if we introduce the CLs after the replication is complete, referred as Arc-2* in Fig.2c, then the (mean) time $\tau_S$ for segregation increases. When chain crossing is not allowed for a simple ring polymer architecture (Arc-0* with 500 monomers) they often get entangled and can only segregate in 40% of the attempts. When topoisomerase effects are included by allowing chain cross as for Arc-2, Arc-2* and Arc-0, the chains segregate about 80% of the cases. However, we checked that shorter ring polymers with 100 monomers can easily segregate even without effects of topoisomerase, as seen previously [17]. The values of $\tau_S$ for different architectures is given Fig.2c; $\tau_s$ is calculated by averaging over the runs that lead to successful chromosome segregation, from a pool of 10 independent runs. Thus we infer the quicker segregation with the Arc-2 polymer is a consequence of the presence of internal loops in each ring-polymer. We show that each loop-COM gets separated along the cylinder-axis once the loops get created by CLs, refer SI-IV Fig.S3.

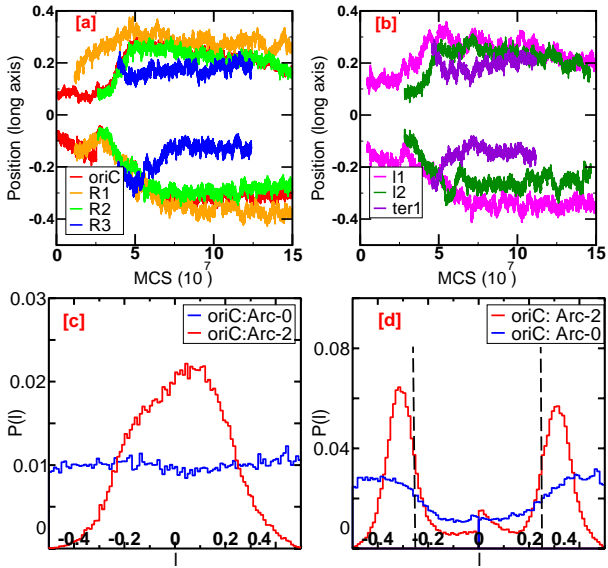

Figure 3. **Dynamics and localisation of chromosomal loci with Arc-2 polymer architecture:** Trajectory of the (a) *oriC*, R1, R2 and R3 loci and (b) L1, L2, and ter1 loci after the replication of the locus, along the cylinder long axis. To plot the locus trajectories, we averaged over 20 indepen- dent runs and scale the position by the final length of the confining cylinder. We follow the conventions of the 'polar orientation' as in [23], where the replicated locus is assigned the negative axis. (c) Comparison of the probability distribu- tion of the *oriC* position along the cylinder long-axis for the parent DNA (before replication) for Arc-0 and Arc-2 polymer architectures. We note that Arc-2 shows a narrow distribu- tion around the midcell while for Arc-0 oriC shows a uniform distribution. (d) Comparison of the probability distribution of the oriC position post replication, for Arc-0 and Arc-2. Data is collected from 20 independent runs from 10 15 $10^7$ MCS. The black dashed lines represent the quarter positions and have been added for aid of visualization. We note that Arc-2 shows a narrower distribution approximately around the quarter positions, as compared to Arc-0.

**Localization of oriC to quarter positions:** At the first step of the replication process of our model, a new *oriC* monomer is created and is named the *oriC* of the D2-chain. We see in Fig.2d, that the two (replicated) *oriC*s are at the middle of the cylinder at the start. The two additional CLs are created after 2.5 $10^7 \times$MCS. Once the CLs are introduced between the newly added 125-th and the 375-th monomer and *oriC* of the DNA-2-chain, we see sharp change in *oriC* position of the DNA-2 poly- mer chain (bottom red line). For the monomers on the other chain, we only change the monomer status from mother to daughter as RF position moves along the con- tour and rename it monomers of DNA-1, and hence the CL between *oriC*-125 and *oriC*-375 monomers existed even at the start of replication. After the loops of DNA-2 are created, we observe that the oriC-s start migrating out to occupy the $1/4$ and $3/4$ positions (scaled by the final length of the cylinder $L_f$) on the elongated cylinder axis.

**Localisation of other genomic loci.** In Fig.3(a,b) we follow the positions of the monomers corresponding to loci labelled as *oriC*, R1, R2, R3, L1, L2 and *dif-ter* [23], as function of MCS. We see that different loci seg- regate out to two different halves of the cylinder after replication; data is plotted from the stage that the RFs reach the individual monomer locus, and is replicated. We observe that though the *oriC* is replicated at the be- ginning of the simulation, it branches out to two halves of the cylinder only after the loops in D2 chain are created. This matches with the observation of cohesion time in experiments. Other than *oriC* and *R*1, other loci have extremely low cohesion times. Data shown is averaged over 20 independent runs, starting from statistically in- dependent initial conditions.

The mutual repulsion between internal loops in an (un- replicated) DNA-polymer with Arc-2 architecture also lo- calizes the *oriC* to the middle of the cylinder at the be- ginning of replication process. We create an equilibrium structure of a single Arc-2 DNA-polymer to simulate an unreplicated chromosome and we observe that the *oriC* localizes to the middle of the cylinder with $L_i = 17.5a$ which is as seen in-vivo: refer Fig.3c. In contrast, the *oriC* of a ring polymer with 500 monomers (Arc-0) is found non-localized along the cylinder axis under equi- librium conditions.

Subsequently, if replication is switched on, the *oriC*s move to the two quarter positions along the cell axis for the (daughter) DNA-polymers with Arc-2, as shown in Fig.2d. In Fig.3d, we plot the probability distribution $P(l)$ of finding the two *oriC*s of DNA-1 and DNA-2, and show the distribution is sharply peaked around the quarter positions for Arc-2 architecture as compared to the two oriCs daughter-polymers with Arc-0 architecture. Similar data for the other loci is given in SI-V and in Fig.S4. Thus the presence of loops localizes all the differ- ent loci in complete agreement with experiments reported in [23]. As we see in Fig.3a,b all loci reach their home positions after segregation by the end of $8 \times 10^7$ MCS. But we start collecting collect statistical data to plot the distributions in Fig.3d after $10 \times 10^7$ MCS from the start of simulation. Thus each loci reaches its home position before we collect statistical data for the next $5 \times 10^7$MCS to calculate $P(l)$. We do not model the division of the cell, hence each locus continues to fluctuate around its home position till $15 \times 10^7$ MCS . We can unambigu- ously conclude from the width of each distribution that each loci is localized at their home positions, and the two *oriC*s, specifically are at the quarter and three-quarters positions, respectively.

**Mechanism of localization of genomic loci:** How does the architecture Arc-2 lead to the localisation of *oriC* to the mid-cell position of an unreplicated chro- mosome and to the quarter positions for two daughter chromosomes? We examine architecture Arc-2 in detail in Fig.4(a). Arc-2 has two side loops of length 125 monomers each and a bigger loop of 250 monomers. In Fig.4b, we plot the probability distributions of the positions of the center of masses (c.o.m) of the individual loops of the daughter chromosomes along the cylindrical axis. We observe that the two daughter DNA-polymers repel each other and thus occupy different cell halves of the cylinder. Due to the modified polymer architecture with internal loops, the two smaller loops (loop-1,2) further repel the bigger loop (loop-3). We further infer that loop-1 & loop-2 overlap and occupy the same volume within the cylinder, even as they repel loop-3. Using the blob picture for polymers, we have checked that an overlap of loop-1 (or loop-2) with loop-3 has a higher entropic cost, we report this in the Supplementary Information of [64].

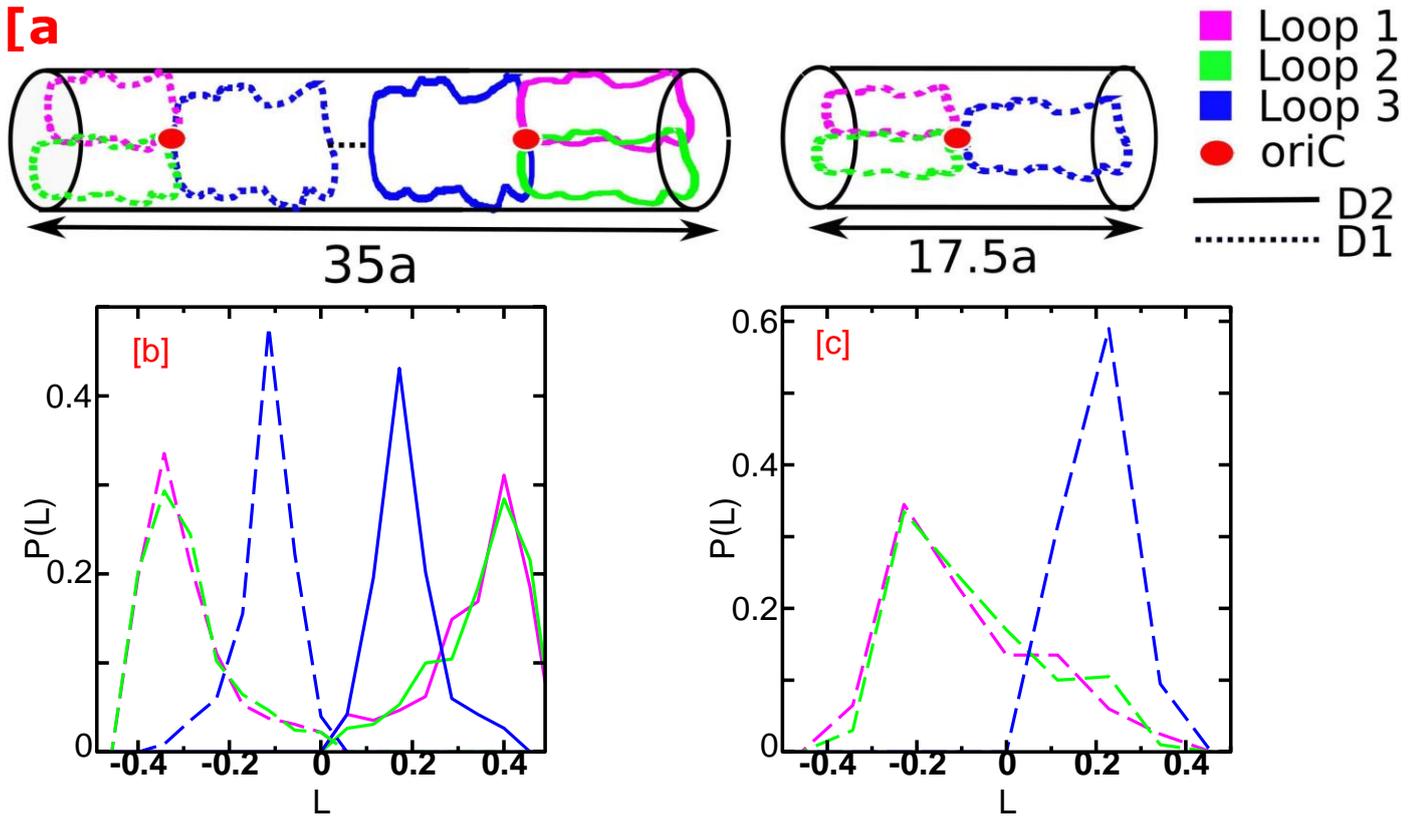

Figure 4. **Mechanisms of *oriC* localisation : Entropic repulsion between loops** (a) and (b) shows a schematic of two daughter chromosomes (linked at the *dif-ter*) having Arc-2 architecture, and a schematic of a single unreplicating mother chromosome (D0) having Arc-2 architecture. Subfigure (b) shows the probability distributions $P(L)$ of the center of masses (c.o.m.) of the three loops for both the daughter chromosome : DNA-1 (D1) and DNA-2 (D2)). The $P(L)$ have been computed from equilibrated (unreplicating) conformations of DNA-1 and DNA-2 confined in cylinder. To plot, we adopt a convention where we plot the distributions of the c.o.m. of the loops of polymer D2 on the negative x-axis while those of D1 have been plotted on the positive x-axis. The x-axis has been scaled by $L_f$. There is minimal overlap in the $P(L)$ of loops-1 & -2 with loop3 (for each of D1 and D2). The locus *oriC* is by design the junction of these 3 loops and hence gets localised to the quarter positions. Subfigure (c) shows the $P(L)$ of the three loops of D0 with the Arc-2 architecture. The relatively lower overlap in the $P(\ell)$ of loops-1 & 2 with the $P(L)$ of loop-3 indicate the entropic repulsion between the smaller and bigger loops. The *oriC* locus is again at the junction of the three loops, which leads to its mid-cell position, as seen in Fig.3c. The other loci also get localised to specific regions in the cylinder (cell). The distributions in (b) and (c) have been computed 50 independent runs.

A similar scenario holds for the parent DNA which pervades the whole of the cylinder of length $L_i$. The $P(L)$ of the loops for the unreplicating mother DNA in presented in Fig.4(c). Since the locus *oriC* is the junction of the three loops by design, the *oriC* locus gets localised to the middle of center of the each of the polymers. Thus for the parent DNA, the oriC is localized to the middle of the cylinder, and to the quarter positions for the cylinder of length $L_f$ with DNA-1 and DNA-2. Furthermore the other loci (belonging to particular loops) also get localised to specific regions of the cell (cylinder).

**Drift velocity of segregation** We measure the drift velocity of segregation of the individual loci after its replication, to establish that we semi-quantitatively reproduce the *in-vivo* dynamics of loci-segregation, as seen in Fig.7 of [23]. Our procedure for calculating the drift velocity is outlined in SI, section VI. We show that we reproduce some aspects of the universal values of velocities observed for all loci. Our calculated values of velocity have very similar values as measured experimentally, 5 minutes after the onset of loci-segregation. We have not looked at smaller time intervals as in [23], because of the difficulty in getting good statistical averages for the value of the speed of certain loci. But more strikingly, we reproduce similar values of drift velocity (within the acceptable range of fluctuations), [23], at similar times after the replication of reference loci, refer Fig.5. However, in [23] the *oriC* locus shows a higher drift velocity than other loci at long times, viz. after 15 minutes after replication. A more detailed model incorporating the mechanism of the approach of of CL-monomers towards each other may reproduce this aspect, as our model is minimalist in nature.

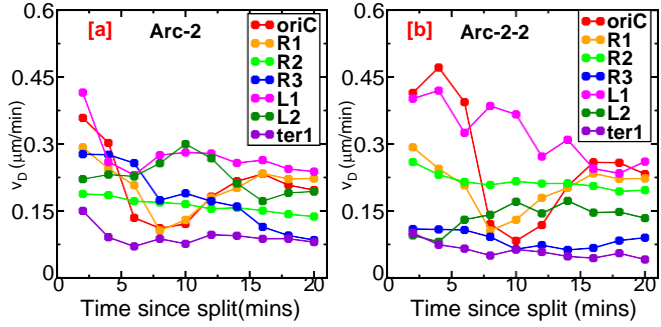

Figure 5. **Drift velocity of loci** The drift velocity $v_D$ for individual loci as a function of time (converted from MCS) obtained with (a) Arc-2 and (b) Arc-2 2. Completion of replication takes 5 $10^7$ iterations or 33 mins for both cases. Details of methodology for calculating $v_D$ is outlined in the SI-VI. The drift velocity of loci show universal behavior, i.e, value of $v_D$ are quite similar close to the onset of segregation, as well as later times, as shown in-vivo [23]. Conversion from MCS to time units enables comparison with experiments[23].

**Macrodomain organisation : Arc-2-2 architecture** We need to use Arc-2-2 architecture instead of Arc-2 architecture to match macrodomain-like organisation of the bacterial chromosome as observed in [25]. Modeling results using Arc-2 architecture could reproduce several experimental results related to spatio-temporal dynamics of DNA loci in the B, C period of *E. coli* cell cycle, but does not produce the coarse-grained Hi-C Map. The Hi-C data is experimentally obtained for a large ensemble of cells, each of which are at different stages in their cell cycle. We show here a simulated contact map using data for the coordinates of different monomers over the entire life cycle i.e. upto $8 \times 10^7$ MCS plotted in genomic coordinates with the Arc-2 architecture (refer Si-VII, Fig.S5). We give more details of how we collect data for Contact maps in the SI-VII.

However, a modification of our polymer architecture by adding two additional CLs, between monomer number

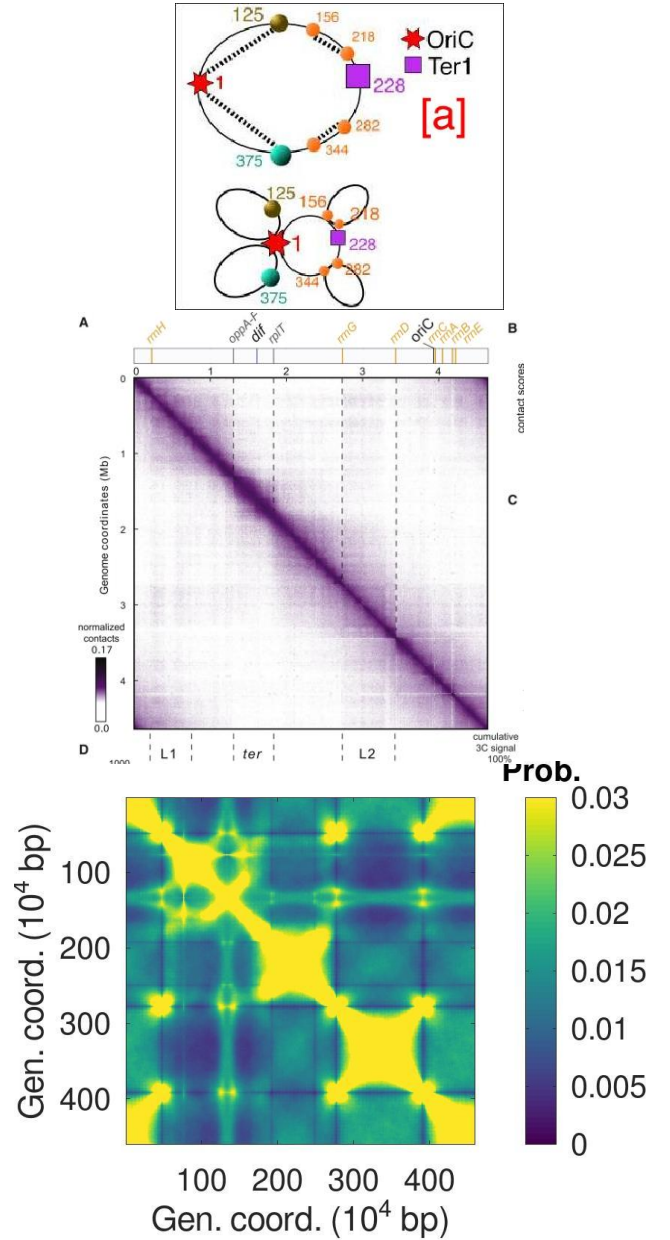

Figure 6. **Macrodomain organisation obtained with polymer architecture: Arc-2-2** (a) Schematic showing the cross-linking sites in Arc-2-2 architecture. Subfigure (b) shows the experimental contact map for E.coli chromosomes as published in [25] (requisite permissions obtained). The fig- ure is being shown for aid of comparison with the simulated contact map. Subfigure (c) shows the contact map obtained from simulations with Arc-2-2 architecture, averaged over10 independent runs.

156 and 218 on one arm and 344 and 282 on the other arm, leads to the Arc-2-2 architecture. This architecture results in a contact map that has 4 bright domains roughly of size 1$Mbp$ each refer Fig.6 and thus leads to a macrodomain-like organization of the chromosome-polymer. Moreover, the Arc-2-2 architecture also reproduces (or betters) all the segregation and localization dynamics discussed till now with the Arc-2 architecture.

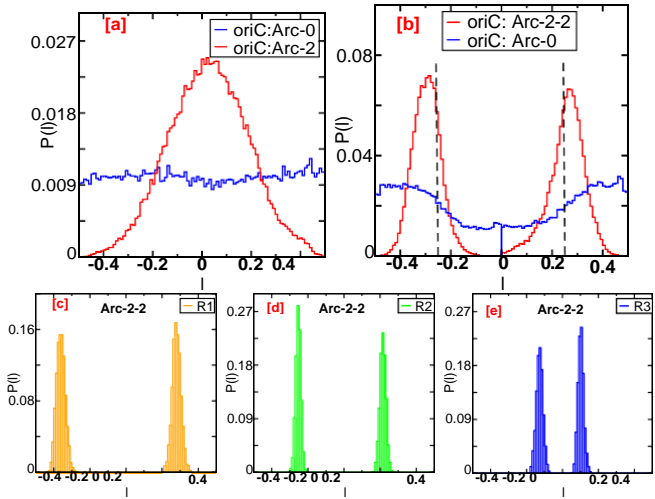

Figure 7. **Localisation of loci: Arc-2-2** (a) Probability distribution $P(l)$ of the oriC position before replication, for Arc0 and Arc-2-2, averaged over 10 independent runs. We note that Arc-2-2 shows a narrow distribution around the mid-cell while for Arc-0 oriC shows a uniform distribution.
(b) $P(l)$ of the ori-C position post replication, for Arc-0 and Arc-2-2. Data is collected from 20 independent runs from 5 to $15 \times 10^7$ MCS. The black dashed lines represent the quarter positions, and are added for aid of visualization. We note that Arc-2-2 shows a narrower $P(l)$ distribution around the quarter positions, as compared to Arc-0. $P(l)$ for the positions of (c) R1, (d) R2 and (e) R3 locus post replication for Arc-2-2. The probability distributions in (c), (d) and (e) have been calculated from the averaged trajectory curves (refer SI-VIII: Fig.S6), from a simulation of 5 to $15 \times 10^7$ MCS.

The first macro-domain (L1) extends from $0.2 - 1.2$ Mbp, the second (ter macrodomain) from approximately $1.2 - 1.8$ Mbp and the third (L2) from $2.8 - 3.5$ Mbp. Thus we note that the positions of our macrodomains are in fair agreement with the experimentally determined contact map given in Fig.6b, which has been reproduced from[25]. Moreover, the Arc-2-2 leads to lower segregation times $\tau_S$ as compared to Arc-2 (Fig.2c), and also localizes the *oriC* and other loci even better than the Arc-2 architecture with very similar migration dynamics : refer data presented in Fig.7 and SI-IX, Fig.S7 in the Supplementary. To show that the obtained results are robust, we validate our results for similar simulation studies but with 1000 monomers, where each bead represents 4.6 kbp of DNA (refer SI-X and Fig.S8 in the Supplementary). The mechanism for the formation of macrodomain-like organisation is the repulsion between entropic loops. It was already hypothesised earlier that formation of 'topological domains' crucially depends on entropic repulsion between loops[67]. We hereby show that this mechanism can reconcile the observations from Hi-C experiments done on *E.coli* chromosomes using our proposed polymer architecture Arc-2-2.

**Polymer Architecture may influence Repliosome dynamics:** We plot the trajectory of the RFs for Arc-2 and Arc-2-2 architectures to understand whether the we can resolve a long standing controversy regarding the spatial positions of the RFs. We find that the trajectory shows an 'M' like pattern, which is directly indicative of the underlying polymer architecture (refer SI-XI and Fig.S9). Thus the replication forks are located close to the mid-cell position both at the beginning, middle and the end of the replication process. There is now increasing evidence that the two replication forks remain spatially proximal despite "transient separations of sister forks during replication" [68]. Thus, it is plausible in our opinion, that a detailed study of the underlying polymer architecture may resolve the differences in the "Train track model" and the "Replication factory' models of replication, and the replication forks may remain spatially proximal despite moving in a 'train track' fashion along the chromosome-polymer contour, as the architecture of the polymer maybe suitably modified above and beyond a simple ring architecture .

## III. DISCUSSION

Past studies have indicated that loops or domains play an important role in the organization of chromosomes into macrodomains [69, 70]. However no previous study to the best of our knowledge, has investigated the role of loops/domains on the linear ordering of genetic loci. We propose that a certain looped architecture not just leads to the linear ordering of loci but also the macrodomain organisation of *E.coli* chromosome. We establish in this manuscript that a modification of the polymer architecture by adding 4 loops (or 2 loops) to the ring-polymer to form the Arc-2-2 architecture (or Arc-2 architecture) is favourable for *E. coli* DNA-polymer efficient segregation during replication. We hypothesize that such "effective" cross-links across large genomic distances may arise due to the presence of MukBeF (or other SMCs) which are known to bridge distant chromosomal regions. The observation of macro-domains in contact maps and the similarity in the values of drift velocity observed for all loci emerge organically as a consequence of the adopting the Arc-2-2 architecture. The reason for the localization of the two *oriC*s (to the quarter positions), and other loci is the entropic repulsion between the loops within each individual daughter chromosome. The loops belonging to the same daughter try to occupy different positions within the two halves of the cylinder.

A previous simulation study [50] has highlighted the role of macrodomain organization in the enhancement of entropic repulsion between daughter chromosomes. In the study certain genetic loci such as the *ori* and the *ter* were constrained and their role on the genome organization was studied. We however take a different approach and do not assume any such organisation of loci *apriori*. We just introduce two or four crosslinks and show that there is a spontaneous emergence of loci organisation, with the organisation patterns being similar to those seen *in-vivo*. Our approach is markedly different from that of previous works such as that of [24] and [50], where a MukBEF gradient or other constraints are imposed to achieve the organisational patterns of loci seen *in-vivo*.

Our study also suggests a potential reconciliation mechanism of the train track model and replication factory models regarding the location of the replication fork. Initially the replication begins at the cell-center as the *oriC* of the unreplicated chromosome is localized there. Then the site of replication migrate out from the cell center as daughter polymer architecture gets modified to Arc-2 or Arc-2-2. But finally replication ends near the cell center as *dif-ter* locus gets localized at cell center midway through the replication. In an earlier simulation study [12], the unreplicated DNA polymer was externally constrained within a smaller cylinder within a concentric shell arrangement. However, we did not have to use the concentric shell model of [12], as compaction of the mother DNA-polymer was achieved to some degree, by the Arc-2-2 (or Arc-2) architecture of the unreplicated chromosome at the beginning of replication. There have been conflicting experimental reports with regards to the organization of *E.coli* chromosomes in slow growth. For instance [71] proposes a 'Sausage' model of organisation of the chromosome whereas [23] supports the 'doughnut' organisation. In our work we obtain the 'doughnut' like organisation of the chromosome as in [23] despite employing a modified ring polymer architecture.

Monte Carlo was the simulation technique of choice as we primarily relied on diffusive dynamics to achieve entropic segregation. It also enabled us to (a) introduce springs to model CLs, (b) change the diameter of the monomers at regular intervals to allow chain crossing and (c) also have fast and many independent runs (as compared to Molecular dynamics). The frequency with which we change monomer diameter to allow for chain crossing is important to achieve successful segregation. If we reduce diameter too frequently or have small diameters for prolonged periods, entropic forces arising from excluded volume interactions will be suppressed. On the other hand, we checked that if the release of topological constraints is too infrequent, that again leads to inefficient segregation. We have a mapping of time scales which enables us to introduce monomers every $2 \times 10^5$ MCS which is very close to experimentally known rates. This mapping relies on the monomer size being $\approx 150$ nm incorporating compaction effects at lower length scales. We report consequences of choice of different replication rates, as well as calculations of free energy differences for systems of polymers (with different architectures) in mixed state versus in segregated state in future communications, along with data for fast growth.

While the match of the calculated contact map with the experimental map is a consequence of the presence of the long-lived links between very specific monomers, our prediction of the location of CLs along the chain contour can now be explicitly verified. We had checked that change in position of CLs by 4-5 monomers on either side of the chain contour does not significantly alter results. A past study has also indicated that nonthermal forces mediated through a vico-elastic medium are also involved in the segregation process [54]. In our opinion such forces, perhaps induced by proteins help the bacterial chromosome to attain modified polymer architectures which assist entropic segregation. However, how the SMC proteins bridges different DNA-segments well separated in space within the cylindrical cell has remained an open question and remains so [32]. Moreover, the cell is quite packed with organelles which act as crowders, which could affect segregation dynamics and organization [72, 73]. The presence of such crowders also leads to a significant compaction of the nucleoid [14]. The bacterial chromosome is also significantly more compacted, however we believe this does not affect the central claim of the work which pertains to the organisation of genomic segments. However, more detailed studies incorporating these aspects may prove to be particularly useful to understand their role in assisting entropic segregation of chromosomes.

## IV. AUTHOR CONTRIBUTIONS

DM and AC designed the work and wrote the manuscript. DM did most of the simulations and analysis presented. SP conducted additional research to strengthen our conclusions.

## V. ACKNOWLEDGEMENTS

A.C., with DST-SERB identification SQUID-1973-AC-4067, acknowledges funding by DST-India, project MTR/2019/000078 and discussions in meetings organized by ICTS, Bangalore, India. A.C. thanks Suckjoon Jun for drawing our attention to some unresolved aspects pertaining to bacterial chromosome segregation for fast growth conditions back in 2018.

# Supplementary: Polymer architecture orchestrates the segregation and spatio-temporal organization of replicating *E. coli* chromosomes in slow growth.

Debarshi Mitra, Shreerang Pande & Apratim Chatterji [*]

*Dept of Physics, IISER-Pune, Pune, India-411008.*
(Dated: December 23, 2021)

The mechanism and driving forces of chromosome segregation in the bacterial cell cycle of *E. coli* is one of the least understood events in its life cycle. Using principles of entropic repulsion between polymer loops confined in a cylinder, we use Monte carlo simulations to show that the segregation dynamics is spontaneously enhanced by the adoption of a certain DNA-polymer architecture as replication progresses. Secondly, the chosen polymer-topology ensures its self-organization along the cell axis while segregation is in progress, such that various polymer-loci get spatially localized. The time evolution of loci positions quantitatively match the corresponding experimentally reported results, including observation of the cohesion time and the ter-transition. Additionally, the contact map generated using our bead-spring model reproduces the four macro-domains of the experimental Hi-C maps. Lastly, the proposed mechanism reproduces the observed universal dynamics as the sister loci separate during segregation. It was already hypothesized and expected that SMC proteins, e.g. Mukbef contribute over and above entropic repulsion between bacterial-DNA ring-polymers to aid the segregation of daughter DNAs in the *E.coli* cell cycle. We propose that cross-links (plausibly by SMC proteins) at crucial positions along the contour is enough to provide sufficient forces for segregation within reasonable time scales. A mapping between Monte Carlo diffusive dynamics time scales and real time units helps us use experimentally relevant numbers for our modeling.

## I. MAPPING SIMULATION TIME TO REAL TIME: AN ESTIMATE

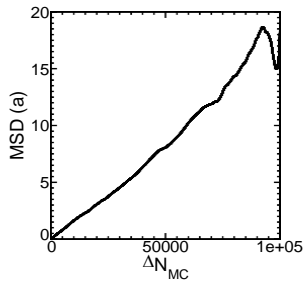

Figure S1. Fig.1 shows the mean square displacement (MSD) of the center of mass of a polymer of 100 monomers as a function of time, in the absence of confinement.

To study phase separation kinetics using Monte Carlo (MC) techniques to model underlying diffusive phenomenon, researchers often express the time evolution of the phenomenon in terms of Monte Carlo iterations cite(binder, Puri). We use Monte Carlo to study segregation dynamics in the same spirit and assume a linear relationship between the elapsed time and the progression of Monte Carlo steps. We further try to map $N_{MC}$ MC iterations to time in real units e.g. 1 second.

To obtain a mapping of our Monte Carlo iterations to real time, we plot the mean square displacement (M.S.D) of the center of mass of a ring polymer of $N_p = 100$ monomer-beads as a function of time, in the absence of confinement as shown in Fig.S1. The unit of length in simulation is $a$, which is the equilibrium length of the spring (bond) between two adjacent monomers. We estimate a displacement of $8a^2$ in $\Delta N_{MC} = 50000$ iterations to obtain a diffusion constant of $D_p = 8a^2/(6\Delta N_{MC})$. We obtain the diffusion constant of a single monomer from our simulations, $D_{sim} = D_p \cdot N_p$, assuming Rouse dynamics [1]. Thus, from Fig.S1 we note that-:

$$D_{sim} = \frac{8a^2 N_p}{6\Delta N_{MC}}. \tag{1}$$

We now estimate $a$ to be approximately 150nm. Each monomer bead in our simulations represents 9.2 kbp. The Kuhn-length of bare DNA is known to be around $\ell_K = 100$nm and contains 300 bp. Thus assuming ideal chain statistics of 9.2 kbp segment equivalent to 30 Kuhn monomers with a corresponding length scale of $\sqrt(30)\, \ell_K$, and about 4 times compaction due to NAPs and other condensation mechanisms such as super coiling etc, each bead containing 9.2kbp would be of size 150 nm $a$. Now if we choose the diameter of our confining cylinder (as in the simulations described in the main text) to be $7a$, then on conversion we get the cell diameter is roughly $1\mu$. Choice of the bead diameter $\sigma = 0.8a$ (which is also essential to prevent crossing of chains) leads to the colloidal volume fraction of 0.2 for a polymer with 500 beads in a cylinder of diameter $D_c = 7a \equiv 1050$nm and length $L_i = 17.5a \equiv 2625$nm.

The estimate of the value of diffusion constant $D_{th}$ of a freely diffusing spherical bead of size $a$ is

$$D_{th}(m^2/s) = k_B T/(6\pi\eta a) \tag{2}$$

where $\eta$ denotes the viscosity of the cellular matrix. Assuming $\eta$ to be that of water and $T$ to be the room temperature, we obtain a correspondence from Eqn.1 and

[*] Correspondence email address: apratim@iiserpune.ac.in

Eqn.2 between simulation time (iterations) and real time. We now equate:

$$D_{sim} = D_{th} \qquad (3)$$

to obtain $1s \approx 25000$ iterations.

We use the above mapping to present MC simulations data in units of real time, i.e. in seconds and minutes.

## II.. SIMULATION MODEL DETAILS

1. In the bead spring model of the ring polymer one considers a chain of spherical particles, where the neighbouring particles are connected by harmonic spring interactions $V_{spring} = \kappa(r - a)^2$, where $\kappa = 100 k_B T / a^2$ is the spring constant of the springs, $r$ is the distance between two neighbouring monomers, and $a$ is the equilibrium length of the springs. For the ring architecture, the 500-th monomer is connected to the 499-th and the 1-st monomer. Chain crossing is not allowed, by having excluded volume interactions between monomers modelled by the WCA potential. Each monomer has diameter $0.8a$ and the expression of the WCA potential is :

$$V_{WCA} = 4\epsilon[(\sigma/r)^{12} - (\sigma/r)^6], \forall r < 2^{1/6}\sigma \qquad (4)$$

where $\epsilon = k_B T$ is chosen. However, chain crossing and release of topological constraints is allowed when the effects of topoisomerase is switched on every 10000 Monte Carlo steps for the next 900 MCS. For 900 MCS, when the effects of topo-isomerase are switched on, one chooses $\sigma = 0.1a$ which allows chains to cross each other.

2. Suppose the two replication forks (RFs) are at position $i$ and $500 - i$. For instance if $i = 5$, i.e the RFs are at the position of the 5-th monomer and the 495-th monomer then in a ring polymer, the 5-th monomer is connected by springs to the 4-th and the 6-th monomer. After replication of the 5-th (and 495-th) monomer, a monomer will be added positionally close to the two RFs, such that the 5-th (and the 495-th) monomer is connected by springs to a third monomer $5'$ (and $495'$). The $5'$-th (and $495'$-th) monomer is in turn is also connected by another spring to the $4'$-th (and $496'$-th) monomer. The RF then moves to the monomer number 6 and 494, respectively, where the above process is repeated after 2 $10^5$ iterations. Refer FIG.S2.

3. We add a monomer at a distance $a$ in a random direction from the replication fork, and link it by springs to (a) bead of the mother DNA at the position of RF, and also (b) to the previously replicated bead of the daughter DNA, refer schematic FIG.S2. Note that RFs are modelled only implicitly, where the monomers are added along the chain contour, in accordance with the train track model of replication.

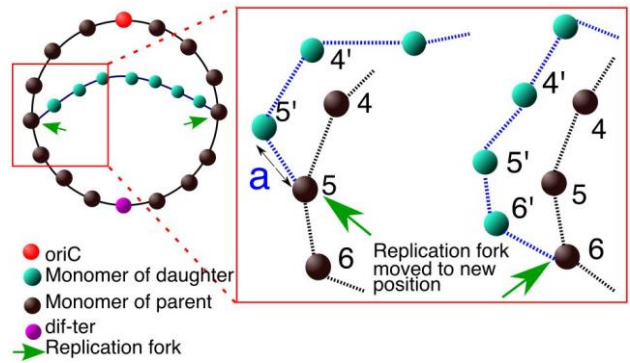

Figure S2. A schematic showing the procedure by which we mimic replication in our simulations. To mimic replication we add a new monomer as the implicit replication forks move (bidirectionally) along the chain contour.

## III.. MEASURING TIMESCALES OF SEGREGATION: PROCEDURE FOR CALCULATION

To measure the time scales of segregation as shown in Fig.2c of the main manuscript we follow the distance between the center of masses of the two daughter chromosomes. If the distance is greater than or equal to $11a$ we consider the chromosomes to have segregated. The corresponding times are treated as the times of segregation. For Arc-0, Arc2, Arc2* and Arc-2-2 , 8 out of 10 independent runs segregate. Thus the error bars have been computed using those "successful" 8 runs. Arc0* however only results in segregation 4 out of 10 independent runs and the error bar therefore has been calculated from those 4 "successful" runs.

## IV.. ENHANCED ENTROPIC REPULSION BETWEEN DAUGHTER CHROMOSOMES DUE TO THE PRESENCE OF LOOPS

Now we wish to understand the mechanism by which the addition of two cross-links enhances the entropic repulsion between the two daughter chromosomes, DNA-1 and DNA-2. We note that these additional cross-links leads to the formation of loops within each daughter chromosome. We name these loops as loop1, loop2 and loop3 for convenience. We remind the reader that the monomers belonging un-replicated Arc-2 polymer is rechristened as daughter DNA-1 after the RF has crossed the corresponding point along the chain contour. Next we follow the z-coordinates of the COM of each of these loops as a function of time. The axis of the cylinder is considered parallel to $\hat{z}$, which is along the cell long axis. These are shown in Fig.S3b. Each color represents a particular loop. Note that there are two lines of the same colour corresponding to each loop. The solid line represents the position of the loops in un-replicated Arc-2 polymer initially, but after bifurcation into two lines, the solid and the dashed lines represent DNA-1 and DNA-2 respectively. The COMs are calculated after all the monomers belonging to a particular loop have been replicated, and CLs have been introduced at the appropriate positions. We note that for the COM for each loop the two lines of the same color, corresponding to daughter chromosomes DNA-1 and DNA-2, drift away from each other as time progresses. This is indicative of entropic repulsion between the old and the newly formed loops. Furthermore the loops loop1,loop2,loop3 of a particular daughter-chromosome also largely occupy different regions along the cell axis and minimize overlaps. The presence of these loops thus enhance the entropic repulsion between the two daughter chromosomes by inducing entropy driven repulsion between the loops belonging to different chromosomes.

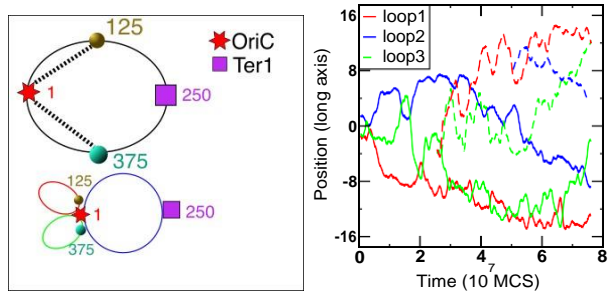

Figure S3. Fig. (a) shows the modified polymer architecture with two additional crosslinks. Note that the addition of these crosslinks results in the formation of loops within each daughter chromosome. Fig.(b) shows the z-coordinates of the COM (center of mass) of the loops (after the constituent monomers have been replicated) within each daughter chromosome for one representative run, running averaged over $10^6$ iterations. The colors in the two subfigures refer to specific loops of each daughter chromosomes. The two lines (one full and the other dashed) corresponding to a particular colour represent the loops of two daughter chromosomes DNA-1 and DNA-2. Note that the dashed blue curve is seen only after $5 \times 10^7$ iterations, as the constituting monomers have only been replicated after $5 \times 10^7$ MCS. The dashed red and the green curves start out from similar times as the constituent monomers get replicated at same times (due to bidirectional motion of RFs). Other independent runs show different dynamics but share similar generic features.

# V.. DISTRIBUTION OF POSITIONS OF LOCI ALONG CYLINDER LONG AXIS POST REPLICATION : ARC-2 ARCHITECTURE

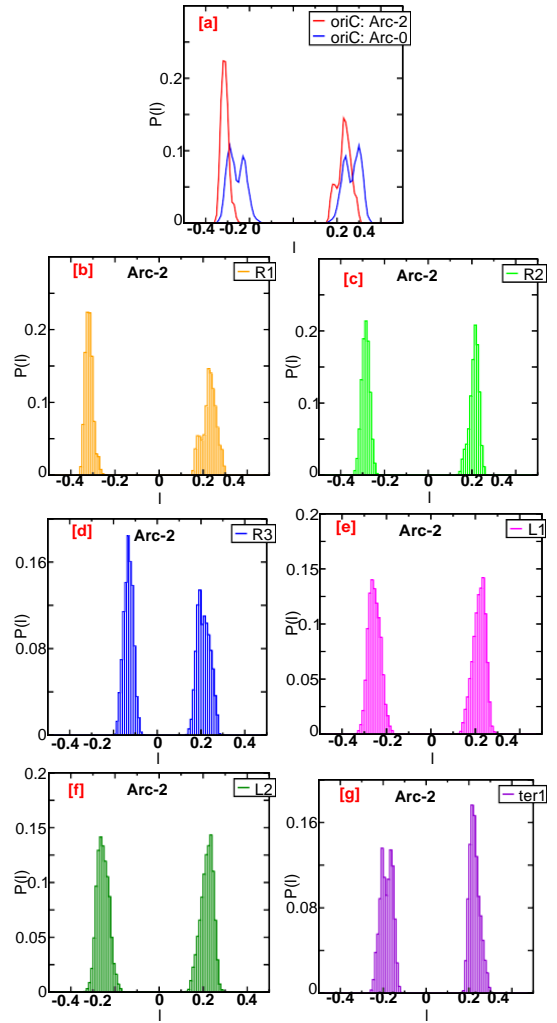

Figure S4. The data shows the probability distribution of the positions corresponding to the (a) oriC (for both Arc-0 and Arc-2 architectures) (b) R1, (c)R2, (d)R3 loci, (e)L1, (f) L2 and (g) ter1 loci along the cylinder long axis for loop architecture Arc-2. The probability distributions in (a), (b), (c), (d), (e) ,(f) and (g) have been calculated from the averaged trajectory curves (given in Fig.3a and Fig.3b of the main manuscript), from a simulation time of $5 \times 10^7$ MCS to $15 \times 10^7$ MCS. Note that the distributions in (a) have lesser standard deviations as compared to Fig.3d of the main manuscript. This is because these probability distributions have been computed from the averaged trajectory curves, which results in a suppression of fluctuations. However we still find that oriC is more sharply peaked around the quarter positions for Arc-2, as compared to Arc-0.

## VI. PROCEDURE FOR MEASURING DRIFT VELOCITY

To measure drift velocity of different loci at different times as the loci segregate, we adopt a similar approach as that of [2], where they obtain the drift velocity at a time $t$ as a finite difference between the positions at time $t$ and $t + \delta t$ where $\delta t = 1$ min, divided by $\delta t$. The symbol $t$ denotes the time since the 'splitting' of the loci. We use the same value of $\delta t$, i.e $\delta t = 1$ min, but we calculate the slope of the trajectory curve of each newly replicated locus (such as those shown in Fig.3a and b of the main manuscript) separated by times $t$ and $t+\delta t$. We define the time of split as the time after which the spatial separation in the segregated and unsegregated locus is greater than $R = 3.5a$ where $R$ denotes the radius of the confining cylinder.

## VII. SIMULATION CONTACT MAPS

We generated a simulated contact map corresponding to Arc-2 in order to explore whether we obtain macrodomain like organisation as seen in-vivo [3]. We note via Fig.S5a that we do not obtain the correct macrodomain organisation as seen in [3]. We then modify the polymer architecture Arc-2 slghtly by adding side loops. This architecture is as shown in Fig.5(a) of the main manuscript. With this modified polymer architecture we obtain Fig.5(c) of the main manuscript. We note that the macrodomain organisation in Fig.5(c) is similar to that of [8]. To generate the contact map we generate a $500 \times 500$ matrix indicating whether any two monomers (genomic segments) are in close proximity to each other. The matrix elements are either 0 (for non-proximal segments) and 1 (for proximal segments). We collect data every 10000 iterations, over a total time interval (in iterations) of $7 \times 10^{7}$ iterations. If any two monomers are closer to each other than a distance of $2a$, we consider them to be in close proximity (contact). A different choice would lead to a less (or more) illuminated contact map, but would not change the spatial organisation of macro-domains. This simulation contact map is then converted to a simulation map in genomic coordinates (as presented here) using the genomic coordinates of loci given in the Supplementary of [2].

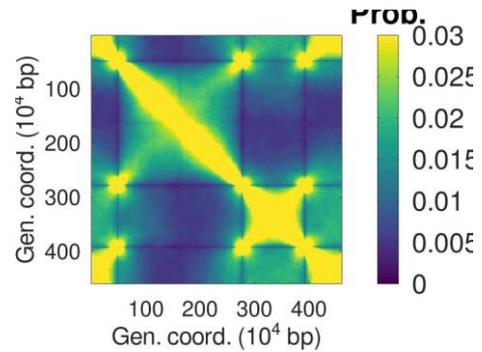

Figure S5. shows the simulated contact map obtained with Arc-2. The organisation obtained is not similar to that obtained by [8].

## VIII. TRAJECTORY CURVES OF DIFFERENT LOCI: ARC-2-2 ARCHITECTURE

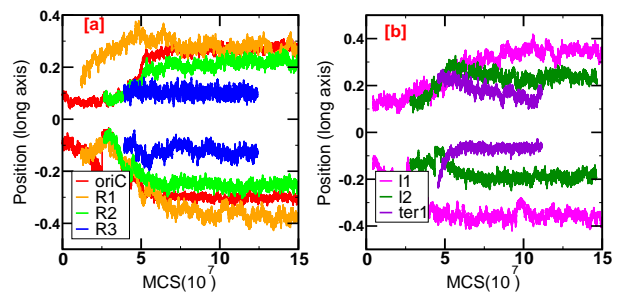

Figure S6. The data shows the positions of the (a) oriC R1, R2 and R3 loci and (b) L1, L2, and ter1 loci post replication, along the cylinder long axis for loop architecture Arc-2 . To plot the locus trajectories we average over 20 independent runs and scale it by the final length of the confining cylinder. We follow the conventions of the 'polar orientation' as in [2], where the replicated locus is assigned the negative axis.

## IX.. DISTRIBUTION OF POSITIONS OF SOME LOCI ALONG CYLINDER LONG AXIS POST REPLICATION: ARC-2-2 ARCHITECTURE

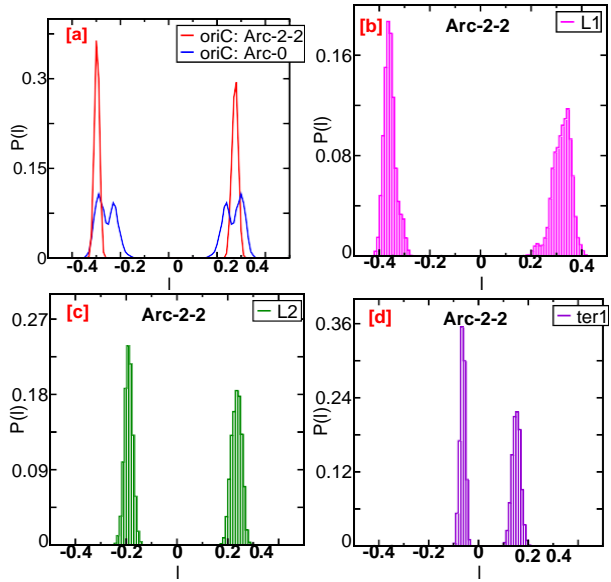

Figure S7. (a)Probability distribution of the oriC locus post replication, for Arc-2-2 architecture. (b)Probability distribution of the L1 locus post replication, for Arc-2-2 architecture. (c)Probability distribution of the L2 locus post replication, for Arc-2-2 architecture.(d)Probability distribution of the ter1 locus post replication, for Arc-2-2 architecture. The probability distributions in (a), (b), (c) and (d) have been calculated from the averaged trajectory curves (given in S.I-VIII and Fig.S6), from a simulation time of $5 \times 10^7$ MCS to $15 \times 10^7$ MCS. Note that the distributions in (a) have lesser standard deviations as compared to Fig.6b of the main manuscript. This is because these probability distributions have been computed from the averaged trajectory curves, which results in a suppression of fluctuations. However we still find that oriC is more sharply peaked around the quarter positions for Arc-2-2, as compared to Arc-0.

the unreplicated and unreplicated chromosomal loci averaged over 10 independent runs. We note that they successfully segregate to different cell halves. Furthermore we show that the locus oriC localises around the quarter positions post replication. Thus our simulation results are robust to variations in the degree of coarse graining.

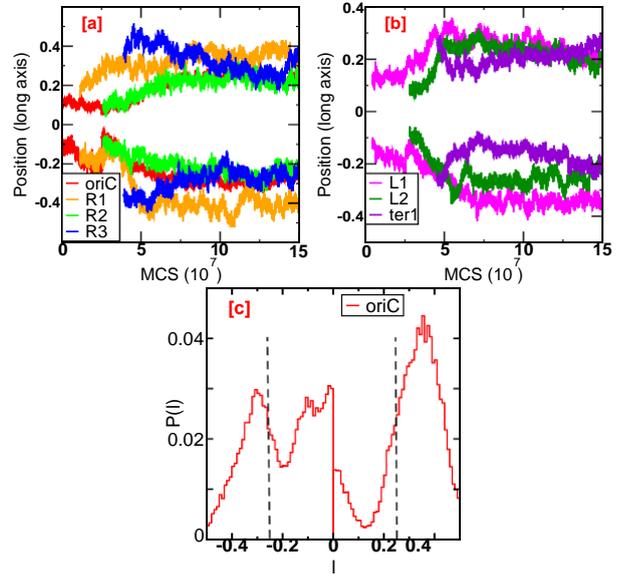

Figure S8. The data shows the positions of the (a) oriC R1, R2 and R3 loci and (b) L1, L2, and ter1 loci post replication, along the cylinder long axis for loop architecture Arc-2, for a polymer chain of 1000 monomers. To plot the locus trajectories we average over 10 independent runs and scale it by the final length of the confining cylinder. We follow the conventions of the 'polar orientation' as in [2], where the newly replicated locus is assigned the negative axis. (c)Probability distribution of the oriC locus post replication, for a chain of 1000 monomers. We note that the two oriCs localise around the quarter positions as seen in-vivo. The probability distribution in (c) has been calculated from 10 independent runs.

## X.. SIMULATION RESULTS FOR 1000 MONOMERS : ARC-2 ARCHITECTURE

**Simulation details:** For a simulation study with 1000 monomers each bead represents 4.6kbp of DNA. We change the dimensions of the confining cylinder such that the colloidal volume fraction is 0.2 and the initial aspect ratio is 1 : 2.5. Each monomer bead is replicated after every 100000 MCS. This ensures that we maintain the same rate of replication as before. The monomer indices of the cross-linking sites have been changed such that similar genomic segments are in contact just as before. The unreplicated chromosome is in the Arc-2 architecture and other details are the same as mentioned before. We show through the following figure the trajectory of

## XI. POSITION OF THE REPLICATION FORKS (RF) FOR ARC-2 AND ARC-2-2

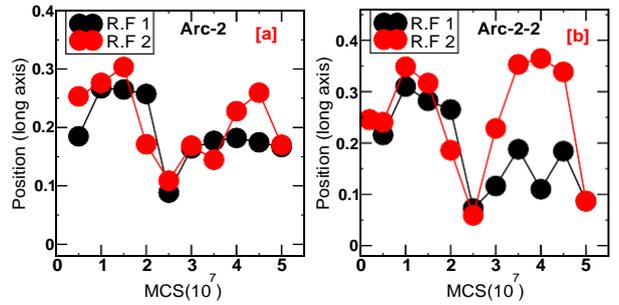

Figure S9. The position of the two replication forks RF-1 and RF-2 as a function of MCS averaged over 10 independent runs, for (a) Arc-2 architecture and (b) Arc-2-2 architecture. The position is scaled by the length of the cylinder $L(t)$ at that particular Monte Carlo step (or equivalently time). The length of cylinder keep increasing incrementally every $3 \times 10^5$ iterations. The range of values on $y$-axis ranges is from from 0 to 0.5 because we take the absolute value of distance of RF-positions from the center of the cylinder.

The replication forks move out bi-directionally from the oriC which is initially localised close to the mid-cell position, for both Arc-2 and Arc-2-2 architectures. Thus in Fig. S9a and (b) we note that the position of the two replication forks are closer to the mid-cell position. The two replication forks then deviate significantly away from the mid-cell position until a simulation time of $2.5 \times 10^7$ MCS is elapsed. It is at this time that the 125-th (and the 375-th) monomers are replicated. Note that for both Arc-2 and Arc-2-2 the 125-th monomer and the 375-th monomer are crosslinked to the oriC, which is localised close to the mid-cell position. Thus at this time the trajectory curves show a sharp dip in the trajectory curve. Only after the replication of the 125-th and the 375-th monomer, does the entropic segregation between the two daughter chromosomes commences. Subsequently the two oriCs start moving to the quarter positions. Thus after $2.5 \times 10^7$ MCS is elapsed, the trajectory curves of the RFs deviate from the mid-cell positions, as they move along the chain contour. The two forks meet at the dif-ter locus which is localised close to the mid-cell position at a simulation time of $5 \times 10^7$ MCS. The trajectory curves of the two R.Fs thus show a 'M' like pattern. This pattern arises solely from the polymer architecture adopted by the chromosomes. Thus it is plausible that the 'Replication factory model' and the 'Train track model' which are seemingly in contradiction to each other, may be reconciled by investigating the underlying polymer architecture.

## XII. MOVIE

We have uploaded a movie file where the unreplicated DNA with 500 monomers is shown in blue inside a cylinder of length $17.5a$ and diameter $7a$. During replication pink monomers of daughter DNA-2 get added and simultaneously the length of the confining cylinder keeps increasing to finally attain a aspect ratio of $1:5$. Finally, we see two segregated DNA-1 and DNA-2 polymers in two halves of the cylinder. The two *oriC*s are shown in red and the *dif-ter* is shown in yellow. The sizes of the *oriC*s and the *dif-ter* locus have been deliberately enlarged for aid of visualization. Note that the polymer architecture changes to the Arc-2 architecture after the replication of the 125th (and simultaneoulsy the 375th) monomer which occurs at ∼$25s$ in the movie. The ter transition (change of ori-ori-ter configuration to the ori-ter-ori) configuration occurs at ∼$47s$. Post the ter transition the *dif-ter* locus continues to fluctuate around the mid-cell position which is as seen in-vivo.